\documentclass[]{aa501}
\usepackage[dvips]{graphicx}
\usepackage[]{amsmath}
\usepackage{psfig}
\usepackage{times}
\newcommand {\chem}[2] {$\rm{}^{#2}\kern-0.8pt#1$}
\def\mabs{$M_{\rm B}$}
\def\hii{H{\sc ii}}

\begin{document}

\title{Chemical evolution of starburst galaxies: \\
       How does star formation proceed?}

\author{  M. Mouhcine\inst{1,2}
        \and
        T. Contini\inst{1,3}  
               }
   \offprints{T. Contini, {\tt contini@ast.obs-mip.fr}}
        \institute{
              Observatoire Astronomique de Strasbourg,
              11, rue de l'Universit\'e, F-67000 Strasbourg, France.
         \and
	 Division of Astronomy \& Astrophysics, University of California, Los Angeles, CA 90095-1562, USA
         \and
             Laboratoire d'Astrophysique de l'Observatoire Midi--Pyr\'en\'ees -- UMR 5572, 14 avenue E. Belin, F-31400 Toulouse, France
                  }   
\date{Received; accepted}
\titlerunning{Star formation history of starburst galaxies}
\authorrunning{M. Mouhcine \& T. Contini}
\abstract{
We compute chemical evolution models to constrain the mode 
and the history of star formation in starburst galaxies as a whole, i.e. 
over a large range of mass and metallicity. To this end, we 
investigate the origin of the dispersion observed in the evolution of 
both nitrogen-to-oxygen abundance ratio and galaxy luminosity as a 
function of metallicity for a large sample of starburst galaxies. 
%
%
We find that the variation of the star formation efficiency, 
in the framework of continuous star formation models, produce 
a scatter equivalent to what is observed in the N/O versus O/H 
diagram for low-mass \hii\ galaxies only. 
However, continous star formation models are unable 
to reproduce i) the scatter observed for massive starburst and UV-selected 
galaxies in the N/O versus O/H relation, and ii) the scatter in the 
\mabs\ versus O/H scaling relation observed for the whole sample 
of starburst galaxies. 
The dispersion associated with the distribution of N/O as a 
function of metallicity, for both low-mass and massive galaxies, 
is well explained in the framework of bursting 
star formation models. It is interpreted as a consequence of 
the time-delay between the ejection of nitrogen and that of oxygen 
into the ISM. These models also reproduce the spread observed in the 
luminosity-metallicity relation. 
Metal-rich spiral galaxies 
differ from metal-poor ones by a higher star formation 
efficiency and starburst frequency. Low-mass galaxies experienced
a few bursts of star formation whereas massive spiral galaxies 
experienced numerous and extended powerful starbursts.
Finally, we confirm previous 
claims (Contini et al. 2002) that UV-selected galaxies are 
observed at a special stage in their evolution. Their low N/O 
abundance ratios with respect to other starburst galaxies is
well explained if they have just undergone a powerful starburst 
which enriched their ISM in oxygen.
}

\maketitle

\section{Introduction}

Retrieving the star formation history of galaxies is essential 
for understanding galaxy formation and evolution. The chemical 
properties of galaxies are closely related to their star formation 
history, and can be considered like fossil records, enabling us 
to track the galaxy formation history up to the present. 

At a given metallicity, the distribution of nitrogen-to-oxygen (N/O) 
abundance ratios shows a large dispersion, 
both at low and high metallicity (Pagel 1985; Coziol et 
al. 1999; Contini et al. 2002). 
Only part of this scatter is due to uncertainties in the abundance 
determinations, and the additional dispersion must therefore be 
accounted for by galaxy evolution models.
Various hypothesis were discussed to be responsible for such a
scatter (e.g., Kobulnicky \& Skillman 1998). 
One scenario invokes the chemical ``pollution'' from the 
N-rich wind of Wolf-Rayet stars within the present starburst 
(Pagel, Terlevich \& Melnick 1986). 
Indeed, small-scale abundance inhomogeneities 
are suspected in a few starburst galaxies (i.e., Walsh \& Roy
1989, 1993; Thuan, Izotov \& Lipovetsky 1996; Kobulnicky et al. 1997). These 
observations suggest that, under some conditions, the metals ejected 
by the massive stars can cool very quickly and pollute the surrounding 
ISM on short ($\sim 10^6$ yr) timescales. However, such localized
enrichment does not seem to occur in most of young starburst galaxies 
(Kobulnicky 1999; Oey \& Schields 2000), even if these objects contain 
numerous Wolf-Rayet stars. 
Morever, Kobulnicky \& Skillman (1998) have shown that the hypothesis 
of localized chemical ``pollution'' cannot explain the 
scatter in N/O at a given metallicity. 
Another explanation for the dispersion in N/O at a 
given metallicity involves differing contributions from primary 
and secondary nitrogen, which essentially amounts to variations 
of the initial mass function (IMF) from galaxy to galaxy. 
Despite some claims of IMF variations with time and environment 
(e.g., Eisenhauer 2001), there is no compelling evidence that the IMF 
varies in local galaxies (Parker \& Garmany 1993; 
Hill, Madore \& Freedman 1994; Hunter et al. 1997). 
Preferential oxygen loss from galaxies with high N/O and more 
effective oxygen retention in galaxies with low N/O may be an 
additional explanation for producing N/O variations at constant 
metallicity. This mechanism has been discussed extensively (e.g.,
Dekel \& Silk 1986; De Young \& Gallagher 1990). It has been invoked 
to avoid the overproduction of oxygen in low-mass galaxies 
(Esteban \& Peimbert 1995), but observational evidence of the 
impact of differential galactic winds on galaxy properties is 
still lacking. 

A natural explanation for the variation of N/O at constant metallicity 
might be a significant time delay between the release of oxygen 
and that of nitrogen into the ISM (e.g. Contini et al. 2002, and 
references therein), while maintaining a universal IMF and standard 
stellar nucleosynthesis. The ``delayed-release'' model assumes that 
star formation is an intermittent process in galaxies 
(Edmunds \& Pagel 1978; Garnett 1990) and predicts that the dispersion 
in N/O is due to the delayed release of nitrogen produced in low-mass 
longer-lived stars, compared to oxygen produced in massive,
short-lived stars. 

There is a lot of observational evidence suggesting that 
the star formation history of galaxies has not been monotonic
with time, but exhibits instead significant fluctuations. 
Galaxies in the Local Group are excellent examples showing a variety of 
star formation histories (see Grebel 2000 for a recent review). 
Further evidence for the ``multiple-burst'' scenario was recently 
found in massive starburst nucleus galaxies 
(Coziol et al. 1999; Schinnerer, Eckart \& Boller 2000; Lan\c con et
al. 2001; Alonso-Herrero et
al. 2001; de Grijs, O'Connell \& Gallagher 2001). Even the low-mass 
and less evolved \hii\ galaxies seem to be formed of age-composite
stellar populations indicating successive bursts of star formation
(Mas-Hesse \& Kunth 1999; Raimann et al. 2000). 

In the framework of hierarchical galaxy formation models, 
large structures such as galaxies grow through the merging process 
of dark matter halos into larger and larger units.
Somerville, Primack \& Faber (2001) have shown that models in 
which burst of star formation are triggered by galaxy-galaxy 
mergers reproduce the observed comoving number density of bright 
high-redshift Lyman-break galaxies. 
Moreover, Kauffmann, Charlot \& Balogh (2001) 
recently explored numerical models of galaxy evolution in which star
formation occurs in two modes: a low-efficiency continuous mode, and 
a high-efficiency mode triggered by interaction with a
satellite. With these assumptions, the star formation history of
low-mass galaxies is characterized by intermittent bursts of star
formation separated by quiescent periods lasting several Gyrs, whereas
massive galaxies are perturbed on time scales of several hundred Myrs
and thus have fluctuating but relatively continuous star formation
histories.

Several chemical models have been developed in order 
to investigate the evolution of the N/O abundance ratio as a 
function of metallicity. However, most of these theoretical 
investigations focused on small-mass dwarf irregular and blue compact 
galaxies (e.g. Matteucci \& Tosi 1985; Pilyugin 1992, 1993; Marconi,
Matteucci \& Tosi 1994; Olofsson 1995; Kobulnicky \& Skillman 1998; 
Bradamante, Matteucci \& D'Ercole 1998; Larsen, Sommer-Larsen \& Pagel 2001).  
Only a few models were dedicated to 
massive spiral galaxies (Diaz \& Tosi 1986; Tosi \& Diaz 1990), and 
nothing has been done to investigate the chemical properties 
of star forming galaxies as a whole, i.e. over a large range of mass 
and metallicity.

The main objective of this paper is to investigate whether low-mass
and massive starburst galaxies have different star formation histories 
using their chemical and photometric properties as observational constraints.
Successful scenarii must be able to reproduce the dispersion for both 
scaling relations of starburst galaxies, namely N/O and absolute magnitude 
as a function of metallicity.

This paper is organized as follow. In Section\,\ref{data} we summarize 
the observational constraints for the chemical properties of 
starburst galaxies. In Section\,\ref{models} we describe the chemical 
and photometric evolution models used to reproduce the properties of 
galaxies, together with the nucleosynthesis prescriptions. The results 
are compared with the observational data and discussed in 
Section\,\ref{res}. Our principal conclusions are summarized in 
Section\,\ref{concl}. 

\section{Observational constraints}
\label{data}


For this work, we consider three samples of starburst galaxies: 
\hii\ galaxies, Starburst Nucleus Galaxies (SBNGs) and a new sample 
of UV-selected galaxies. 
\hii\ galaxies are mostly small-mass and metal-poor galaxies
whereas SBNGs are more massive and metal-rich (see Coziol et al. 1999 
for the dichotomy). The \hii\ galaxy sample is a compilation of irregular 
and blue compact dwarf galaxy samples from Kobulnicky \& Skillman (1996) 
and Izotov \& Thuan (1999). The SBNG sample merges an optically selected
sample (Contini, Consid\`ere \& Davoust 1998; Consid\`ere et al. 2000) and a
far-infrared selected sample (Veilleux et al. 1995). 
The sample of UV-selected starburst galaxies (Contini et al. 2002) 
spans a wide range of oxygen abundances, from $\sim$ 0.1 to 1 
Z$_{\odot}$. These objects are thus intermediate between 
low-mass \hii\ galaxies and massive SBNGs. 

The behavior of these starburst galaxies in the N/O vs. O/H
plane (see Fig.\,\ref{no_oh}) has already been investigated by 
Contini et al. (2002).
At a given metallicity, the majority of UV-selected galaxies 
has low N/O abundance ratios whereas SBNGs show an excess of nitrogen 
abundance when compared to \hii\ regions in the disk of normal
galaxies (see also Coziol et al. 1999). The interpretation of these 
behaviors is not straightforward. 
Possible interpretations of the location of UV-selected galaxies and
SBNGs in the N/O vs. O/H relation could be that UV galaxies are picked 
out at the end of a short episode of star formation following a rather 
long and quiescent period (Contini et al. 2002), whereas SBNGs 
experienced successive
starbursts over the last Gyrs to produce the observed nitrogen
abundance excess (e.g., Coziol et al. 1999) 

The behavior of starburst galaxies in the luminosity--metallicity 
plane (see Fig.\,\ref{magB_metal}) has also been
investigated by Contini et al. (2002). UV-selected and \hii\ galaxies 
systematically deviate from the metallicity-luminosity relation followed
by local ``normal'' galaxies, i.e. without active star formation. 
They appear to be $2-3$ mag brighter than ``normal'' galaxies of similar 
metallicity,
as might be expected if a strong starburst had temporarily lowered 
their mass-to-light ratios. Luminous (\mabs\ $\sim -20$)
UV-selected galaxies behave like massive SBNGs, which show a
significant departure from the metallicity-luminosity relation: they 
have higher metallicities than expected for their absolute $B$-band 
magnitudes. 
This behavior could be understood in the context of hierarchical 
galaxy formation (see Coziol et al. 1998). According to this 
scenario, galaxy bulges form first through violent mergers 
and disks form later, through accretion of residual outlying gas
and/or small gas-rich galaxies. Following Struck-Marcell's (1981) 
models, accretion of more gas than stars will result in a steepening 
of the metallicity-luminosity relation, explaining the behavior 
of SBNGs and massive galaxies in general.

\section{Chemical evolution models}
\label{models}

\subsection{Assumptions} 

The chemical evolution is calculated by solving the usual set of
differential equations for the gas mass density, chemical 
elements mass density and the total metallicity (e.g. Tinsley 1980). 
The model is fully described in Mouhcine \& Lan\c{c}on (2001a). 
We will recall here the relevant parameters of the model.

The evolution of the abundance X$_{i}$ of element $i$ in the 
interstellar medium obeys:  

\begin{equation}
\frac{d(X_i\,M_{gas})}{dt}=-X_i\,\psi(t)+R_{i}(t)+X_{i,f}\,f(t)
\end{equation}

\noindent
where $\psi(t)$ is the star formation rate, $M_{gas}$ is the mass 
of the interstellar gas, $f(t)$ is the gas infall rate, and X$_{i,f}$ 
is the abundance of the infalling gas.
R$_{i}$ describes the rate of gas ejected from stars and returning 
into the ISM. It includes the enrichment from both single stars and 
close binary systems. Type Ia supernovae (hereafter SNe Ia) are assumed 
to originate in close binary systems in which one star at least is a 
C-O white dwarf (Whelan \& Iben 1973), within the range of the binary 
masses $3\,$M$_{\odot}$ and $16\,$M$_{\odot}$. 
The infall of gas from the companion pushes 
the mass about Chandrasekhar limit, triggering a deflagration with 
subsequent disruption of the star. SNe Ia sould be considered since 
they contribute a large fraction of iron.
The ejected stellar elements are assumed to be ejected at the end 
of the stellar life. Dynamical effects are not included in our 
models, we can not account for the internal structure of the ISM. 
We have to assume that the gas is always well mixed within the ISM.     
No instantaneous recycling is assumed which means that both the 
actual lifetime of each single star and the amount of ejected matter 
at that time are considered. 
Galactic winds are neglected and the whole galaxy is treated as 
a single unit from the beginning of its evolution. A one-phase 
interstellar medium is assumed.
The lifetime and ejection rate of chemical elements is strongly 
dependent on the metallicity. In particular, the delayed enrichment 
by Sne Ia is taken into account following the 
prescriptions of Greggio \& Renzini (1983).

The star formation rate is generally believed to be continuous 
with time. The most commonly used relation is a power-law, the 
so-called Schmidt law (Schmidt 1959). It assumes that the large-scale 
star formation rate scales with the gas density of the ISM 
($\psi\,\propto\,M_{gas}^k$), with the star formation efficiency 
as a free parameter. The efficiency is chosen such as to mimic the 
star formation histories along the Hubble sequence. In the litterature, 
the value adopted for the exponent $k$ varies between $k=1$ and $2$.
Recent observations by Kennicutt (1998) of the correlation between 
the average star formation rate and surface densities in star-forming 
galaxies point towards an exponent of $\sim\,1.5$ in the Schmidt 
law. As long as such a law is assumed and the gas infall occurs 
continuously as a smooth function of time, the predicted star 
formation is also a smooth function of time. 

The galaxy is assumed to be built by accretion of gas from the 
intergalactic medium. An exponential infall rate is assumed

\begin{equation}
f(t)=M_{tot}\,\frac{\exp(-t/\tau_{f})}{\tau_{f}}
\end{equation}

\noindent
(Lacey \& Fall 1985) where $\tau_{f}$ is the gas accretion time scale 
for the galaxy formation, and $M_{\rm tot}$ is the total gas mass that can 
be accreted by the galaxy (i.e., $\int_{0}^{\infty}\,f(t)\,dt\,=\,M_{tot}$). 
Note that the adopted analytical formulae for the gas infall is 
arbitrary and that the most critical free parameter is the time scale 
of the gas infall. This expression is intended as a simple model with 
reasonable properties: a high rate at early stages of galaxy 
formation and a lower one at later stages. The gas accretion time 
scale has been assumed to be the same for all galaxy models. 
We assume that the accreted gas is primordial. The primordial 
gas abundances are taken from Walker et al. (1991).

We computed three types of models: models with 
continuous star formation in a closed box, models with continuous 
star formation and gas infall, and models assuming bursting star 
formation and gas infall.

The adopted star formation rate follows the Schmidt law for models
with continuous star formation. When bursting models are considered, 
the star formation rate is given by:

\begin{equation}
\psi(t)=
\left\{
\begin{array}{ccc}
\nu_{s}\,M_{gas}^{k}(t) & \mbox{if} & t_{j}\,\leq\,t\,<\,t_{j}+\tau_{B} \\
0                       & \mbox{if} & t_{j}+\tau_{B}\,\leq\,t\,\leq\,t_{j+1} 
\end{array}\right.
\end{equation}

\noindent
with $t_{j+1}\,=\,t_{j}+\tau_{B}+\tau_{IB}$, where $t_{j}$ and 
$t_{j+1}$ are the j$^{\rm th}$ and (j+1)$^{\rm th}$ burst starting 
time respectively, $\tau_{B}$ is the burst duration, and $\tau_{IB}$ 
is the interburst period. 
$\nu_{s}$ is the star formation efficiency (expressed in unit of 
Gyr$^{-1}$), and represents the inverse of the star formation 
timescale, namely the time necessary to consume all 
the available gas in the star-forming region. 
Each model is characterized by a fixed set of parameters 
(i.e., $T_{gal}$, $\tau_{B}$, $\nu_{s}$, $\tau_{IB}$, 
$\tau_{f}$, $k$), where $T_{gal}$ is the present-day age of the 
galaxy. In the rest of the paper we assume $k=1.5$, and 
$T_{gal}=15$ Gyr.

Once the chemical evolution of a galaxy has been calculated, 
the evolution of its spectrophotometric properties can also be 
followed. To achieve this goal, population synthesis models of 
Mouhcine \& Lan\c{c}on (2001b) are used. The reader is referred to 
this paper for more details. These models span the range 
of metallicities $1/50\le Z/Z_{\odot}\le 2.5$, and include 
all phases of stellar evolution relevant to the photometric 
evolution of galaxies: from zero-age main sequence to supernova 
explosion for massive stars, or to the end of the AGB phase for 
intermediate- and low-mass stars. Pre-main sequence and 
post-AGB are not accounted for in our population synthesis 
model as their contribution to the stellar population light 
budget and chemical enrichment is negligible.
The evolutionary tracks up to the end of the early-AGB stars are 
those of the Padova group (Bressan et al. 1993; Fagotto et al. 1994a,b). 
The extension to the end of the TP-AGB 
phase is constructed using AGB synthetic evolution models.
The stellar spectral library of Lejeune et al. (1997) is used to 
achieve transformation of the theoretical quantities into 
observational ones. Nebular continuum emission is also included in 
the models. The nebular continuum emission coefficients in the 
infrared are taken from Ferland (1980) for HI and He\,{\sc ii}. 
Two-photon emission coefficients are taken from Brown \& Mathews (1970). 
The number of ionizing photons is a fraction $\beta$ of the number of 
Lyman continuum photons computed from our spectral library, while 
the rest is assumed to be absorbed by dust. We take $\beta=0.7$ as 
suggested from the measurements of \hii\ regions in the LMC
(DeGioia-Eastwood 1992).

\subsection{Stellar yields}

In this section we discuss the nucleosynthesis of heavy elements 
which are relevant to this study, namely nitrogen and oxygen. 
Furthermore, we state clearly at the outset that throughout our 
discussion we are considering only the most abundant isotope of 
each of these elements, i.e., \chem{N}{14}, and \chem{O}{16}.

\chem{N}{14} is a key element to understand the evolution of 
galaxies that have experienced successive starbursts
since it needs relatively long time-scales ($\sim 300$ Myr) 
 to be ejected into the ISM. 
The origin of nitrogen has been a subject of debate for some 
years (Vila-Costas \& Edmunds 1993). The basic nucleosynthesis 
process is well understood. \chem{N}{14} is produced mainly in 
the six steps of the CN branch of the CNO cycles within hydrogen 
burning stellar zones (see Cowley 1995 for a nucleosynthesis review). 
Nevertheless the type of star, with respect to mass, that dominate 
\chem{N}{14} production is not
very well defined yet. Henry, Edmunds \& K\"oppen (2000) have shown 
that intermediate-mass stars may be the main producer of \chem{N}{14}.
Indeed, stars with initial masses in the range of 
$3.5-4 M_{\odot}\la M_{\rm init}\la 6-8 M_{\odot}$ 
(the upper limit is depending on details of the stellar 
evolution models) produce secondary \chem{N}{14} when the CN 
cycle operates in the bottom of their convective envelope for each 
third dredge-up event during the thermally pulsating AGB phase 
(Renzini \& Voli 1981; van den Hoek \& Groenewegen 1997; 
Marigo 2001; Mouhcine, in preparation).
For these stars, models predict that the \chem{N}{14} production 
is increasing with metallicity. 
For low-mass stars ($0.8\,M_{\odot}\la\,M_{\rm init}\la\,3-3.5\,M_{\odot}$), 
where no burning is operating at the bottom of their envelope, the
ejecta is dominated by hydrogen and oxygen. However, AGB 
evolution models show clearly that intermediate-mass stars are 
not significant producers of \chem{O}{16}.

Regarding massive stars, it is well known that standard evolution 
models do not produce primary \chem{N}{14} (Woosley \& Weaver 1995).
However recent models taking into account the effect of rotation
on the transport of chemical elements in the evolution of stellar
interiors have revealed mechanisms for production of primary 
\chem{N}{14}. The helium convective shell would penetrate into the 
hydrogen layers, and consequently production of primary \chem{N}{14} 
may be efficient (Maeder 2000).
Woosley \& Weaver (1995) predict higher \chem{N}{14} production than 
Nomoto et al. (1997) at solar metallicity, and indicate that 
\chem{N}{14} production increases with metallicity. 
Note that Maeder (1992) did not include the contribution from 
supernova ejection, only from stellar winds, in his derivation of 
nitrogen yields. 

Maeder (1992) predicts a sizable decrease in \chem{O}{16} production 
with metallicity as the result of a mass-loss process which is 
metallicity sensitive (i.e., $\dot{M}\propto Z^{0.5}$), while 
Woosley \& Weaver (1995) predict that they correlate directly with 
metallicity. Carigi (2000) have shown that the dependence 
of mass-loss on the metallicity for massive stars is crucial to match 
the observed behavior C/O radial distribution in the local Galactic
disk, and hence she found that 
Maeder (1992) yields are more consistent with the data than 
Woosley \& Weaver (1995) or Nomoto et al. (1997) stellar yields 
for the same stars.

For the chemical evolution models, we adopt the nucleosynthesis 
prescriptions from Maeder (1992) for massive stars, and 
from van den Hoek \& Groenewegen (1997) for low- and intermediate-mass
stars. For explosive nucleosynthesis products,
we adopt the prescriptions by Thielemann et al (1986, model W7) 
for type Ia SNe, which we assume to originate from C-O
white dwarfs in binary systems. This yield is assumed to be 
metallicity-independent, which is a reasonable assumption based upon 
the similarity of Thielemann et al's (1986) $Z=Z_{\odot}$ and $Z=0$ 
models. 
The individual stellar lifetimes have been taken from evolutionary 
tracks of Padova group. 
Note that overshooting was used in the calculation of these tracks, 
meaning that the critical initial mass for supernova explosion will 
decrease compared to models which neglect overshooting.
Finally, the IMF is truncated at 0.1 M$_{\odot}$ and 120 M$_{\odot}$. 
In all our modeling, stars are distributed according to a Salpeter 
IMF (1955; $\phi(m_i)\,\approx\,m_{i}^{-2.35}$). 

\section{Results}
\label{res}

In order to understand the observed evolution and scatter in the 
N/O and luminosity vs. metallicity relations described in 
Section\,\ref{data}, we have computed several models considering 
both continuous and intermittent star formation histories.
In our chemical evolution model, we decided to include only the well 
constrained ingredients in order to keep the number of free 
parameters as small as possible, to be confident in our interpretations.  
We assumed that the age zero for a galaxy corresponds to the moment 
when the first burst is occurring.

\subsection{The N/O vs. O/H relation}

\begin{figure*}[!ht]
\includegraphics[clip=,angle=0,width=9cm]{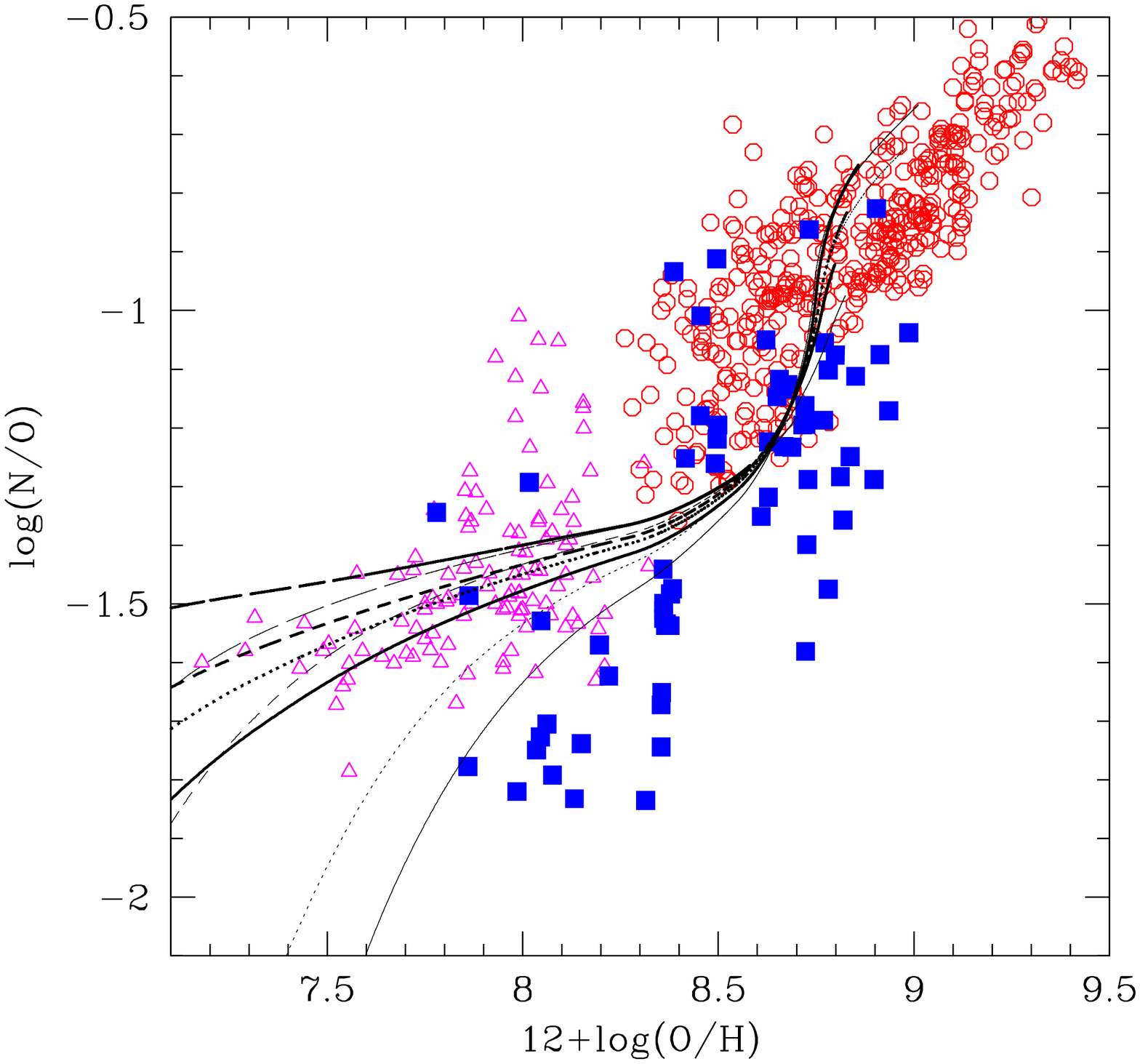}
\includegraphics[clip=,angle=0,width=9cm]{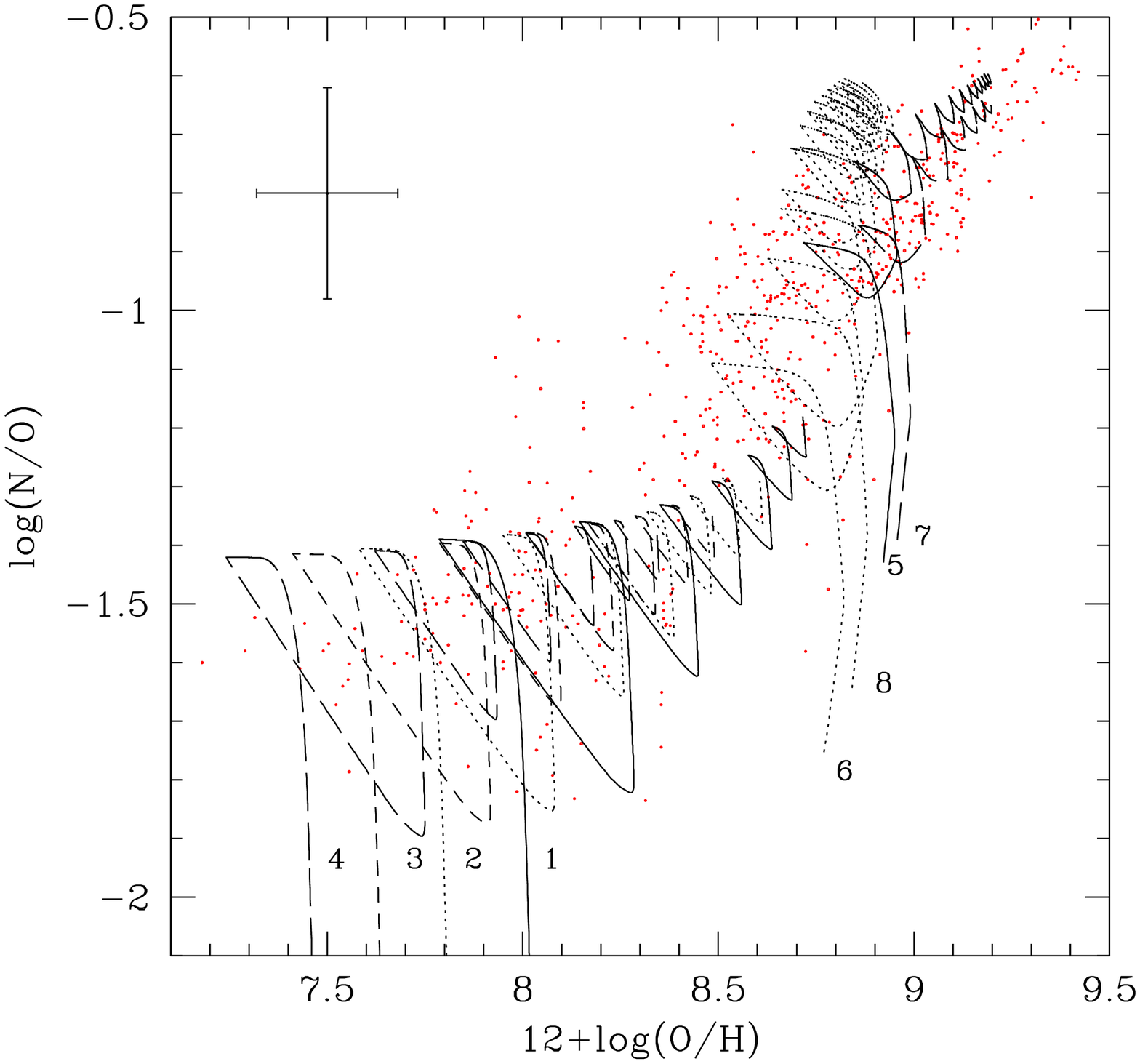}
\caption{N/O vs. O/H for a sample of UV-selected galaxies (squares), 
SBNGs (circles) selected in the optical or in the infrared, and 
\hii\ galaxies (triangles). Typical uncertainties are shown in 
the upper left of the right panel.
The left panel shows model predictions assuming 
a continuous star formation scenario with different star formation
timescales $\tau_{SF}$ (solid line: $\tau_{SF}=2$ Gyr, 
dotted line: $\tau_{SF}=3.5$ Gyr, short-dashed line: 
$\tau_{SF}=5$ Gyr, long-dashed line: $\tau_{SF}=20$ Gyr). 
The thick lines refer to models assuming an infall timescale
$\tau_{f}=7$ Gyr, while thin lines refer to models with 
$\tau_{f}=4$ Gyr. The right panel shows model predictions assuming 
a bursting star formation scenario. The model parameters are listed 
in Table 1. This figure shows clearly the dichotomy between 
the models which reproduce the scatter in the metal-poor region
($12+\log(O/H)\le 8.5$), and those which reproduce the scatter
in the metal-rich region ($12+\log(O/H)\ge 8.5$).}
\label{no_oh}
\end{figure*}

Figure\,\ref{no_oh} (left panel) shows the distribution of N/O as 
a function of metallicity with the corresponding model predictions assuming 
continuous star formation histories with a star formation timescale 
ranging from 2\,Gyr to 20\,Gyr, and assuming that the galaxy forms via gas
infall with timescale on the order of $\tau_{f}=1-7$ Gyr.

Chemical evolution models, assuming continuous star formation, 
are able to reproduce the general trend of the N/O vs. O/H relation. 
Different star formation efficiencies, typical of galaxies along the 
Hubble sequence (from irregulars to early-type spirals; 
Sandage 1986; Fioc \& Rocca-Volmerange 1997; Lindner et al. 1999), 
are shown with different curves in Fig.\,\ref{no_oh}. 
The general behavior of different models (i.e. using different 
SF efficiencies) is the same; the curves are simply shifted toward
higher metallicities for higher star formation efficiencies. 
As expected, the models enter the plot at low metallicity values 
when the efficiency is low, i.e. a star formation efficiency 
typical of irregulars. 

The N/O ratio rises steeply in the high-metallicity regime 
($12+\log(O/H)\ga 8.5$), 
because nitrogen synthesis in intermediate-mass stars is increasing with 
metallicity, while oxygen production in massive stars is decreasing.
In this metallicity range, the N/O abundance ratio 
evolves more rapidly for high values of the star formation efficiency 
than for low ones. This behavior is related to the facts that i) 
the oxygen yield is decreasing with metallicity, and ii) 
the dilution of heavy elements due to gas infall is more efficient 
at high metallicity, for the simple reason that models calculated with a 
high star formation efficiency are metal-rich at younger ages.   
Increasing the gas infall rate affects the evolution of the N/O ratio in 
the sense that the same N/O value is achieved at lower metallicity for 
higher gas infall rate at a fixed star formation efficiency. 

The models indicate that the behavior of low-metallicity galaxies 
($12+\log(O/H)\la 8.0$), located in the region where N/O is nearly 
constant with O/H, is naturally explained if these objects are 
characterized by low SFRs with a large fraction of N 
being produced by intermediate-mass stars (see also Legrand 2000; 
Henry et al. 2000). In this case, we do not need to assume that these 
galaxies are forming their first generation of stars.
Using closed-box models (i.e. without gas infall), the evolution 
of N/O as a function of metallicity is qualitatively similar to 
model predictions with gas infall, except that the predictions 
are scaled-up as no dilution by metal-free gas is operating.

Figure\,\ref{no_oh} (left panel) shows that, assuming a continuous 
star formation 
history, the systematic variation of the model 
parameters (i.e. star formation and gas infall timescales) may 
be a source of scatter in the N/O vs. O/H relation and could thus 
explain the behavior of low-mass \hii\ galaxies. 
However, even considering a large range of free parameters, the 
predicted spread is much less than what is observed for 
metal-rich (i.e., $12+\log(O/H)\ga 8.5$) and UV-selected galaxies.

Theoritical predictions of selected models assuming a bursting star 
formation are shown in 
Fig.\,\ref{no_oh} (right panel). Models parameters are summarized 
in Table\,\ref{parameter}. This plot shows the oscillating behavior 
of the N/O ratio due to the alternating bursting and quiescent phases. 
In this case, the observed dispersion in the N/O vs. O/H relation is 
explained by the time delay between the release of oxygen by massive 
stars into the ISM and that of nitrogen by intermediate-mass stars. 
During the starburst events, as massive stars dominate the chemical 
enrichment, the galaxy moves towards the lower right part of the
diagram. During the interburst period, when no star formation is 
occurring, the release of N by low and intermediate-mass stars occurs 
a few hundred Myrs after the end of the burst and increases N/O at 
constant O/H. The dilution of interstellar gas by the newly accreted 
intergalactic gas is also observed during the quiescent phases.
 
\begin{table}
\centering
\caption{Free parameters of selected bursting star formation models 
shown in Fig.\,\ref{no_oh} (right panel). $\nu_{s}=$ star formation 
efficiency, $\tau_{B}=$ starburst duration, $\tau_{IB}=$ inter-burst 
period, and $\tau_{f}=$ gas infall timescale. See text for more 
details. }
\begin{tabular}{lcccccccc}
\hline
\hline
\multicolumn{1}{c}{{}}&
\multicolumn{1}{c}{{\#}}&
\multicolumn{1}{c}{{$\nu_{s}$ (Gyr$^{-1}$)}}&
\multicolumn{1}{c}{{$\tau_{B}$ (Myr)}}&
\multicolumn{1}{c}{{$\tau_{IB}$ (Gyr)}}&
\multicolumn{1}{c}{{$\tau_{f}$} (Gyr)}\\ 
\hline
Region I  & 1 & 5  & 50 & 0.5 & 4  \\
          & 2 & 3  & 50 & 0.5 & 4  \\
          & 3 & 2  & 50 & 0.5 & 4  \\
          & 4 & 1  & 50 & 0.5 & 4  \\
\hline
Region II & 5 & 25 & 100 & 1.0 & 1 \\
          & 6 & 25 & 100 & 1.0 & 7 \\
          & 7 & 25 & 100 & 2.0 & 1 \\
          & 8 & 25 & 100 & 2.0 & 7 \\
\hline
\hline
\end{tabular}
\label{parameter}
\end{table}

Model predictions with a bursting star formation were thus calculated 
with different sets of the model parameters.
The burst duration was taken to vary from 10\,Myr to 100\,Myr. 
The interburst period varies from 0.5\,Gyr to 7\,Gyr, and the star 
formation efficiency varies from 0.5\,Gyr$^{-1}$ to 30\,Gyr$^{-1}$. 
The number of bursts for each model is computed assuming that the 
galaxy end up with an age of 15\,Gyr. The infall timescale varies 
from 1\,Gyr to 7\,Gyr.
The difference between the models resides in the amount of massive 
stars formed during the burst which is increasing with increasing 
star formation efficiency, increasing burst duration, and decreasing 
gas infall timescale. 

{\it One of the main goals of this study is to determine the 
range of values for the relevant model parameters
that are able to reproduce both the mean evolution 
of N/O as a function of metallicity and its dispersion, and how 
these parameters depend on the galaxy type.}
 
The observed spread of N/O abundance 
ratio is quite well reproduced with the sets of parameters reported in 
Table\,\ref{parameter}.
Extensive model computations show clearly that there is no possible 
combination of the model parameters (i.e., $\tau_{B}$, $\nu_{s}$, 
$\tau_{IB}$, $\tau_{f}$) which would be able to account for 
the observed spread for the {\it whole sample} of galaxies. 
We found that the most important parameter for reproducing the 
observed spread is the star formation efficiency. Once the star formation 
efficiency is set, the extent of the predicted spread mainly depends 
on the starburst duration. The models show that to account for the observed 
scatter of N/O for the whole metallicity range, one needs to consider 
at least two {\it sets} of models characterized by different star 
formation efficiencies and starburst frequency (i.e. number of star 
formation events).

A closer look at the N/O vs. O/H diagram shows that there are two regions
where the N/O abundance ratio evolves differently as a function of 
metallicity. The first region (hereafter region I) is occupied by 
galaxies with $12+\log\,(O/H)\,\la\,8.5$. Those galaxies are mostly 
dwarf irregulars or metal-poor UV-selected galaxies. No correlation 
between N/O and O/H is seen for these galaxies. 
The second region (hereafter region II) is occupied by galaxies with 
$12+\log\,(O/H)\,\ga\,8.5$. Most of them are SBNGs and metal-rich 
UV-selected galaxies. In this region and in contrast to region I, 
galaxies with higher metallicity have also higher N/O ratio.

To reproduce the scatter observed in region I, one needs models 
assuming a relatively low star formation efficiency, between 0.5 and 
5\,Gyr$^{-1}$. Burst durations on the order of $\sim 10-100$ Myr are 
compatible with observations. 
Varying other free parameters has no significant effect on the model 
predictions. These models suggest that galaxies located in region I 
have experienced a few bursts ($n_{b}\sim 1-4$) only. 
When these models enter into region II, they continue to show an 
oscillating behavior, but they produce lower scatter 
in the N/O vs. O/H diagram and have 
steeper evolution than what is observed in region II.

The second set of models, used to reproduce the scatter 
associated to region II, needs a much higher star formation 
efficiency (between 7 and 30 Gyr$^{-1}$) than what is needed to 
reproduce the scatter in region I. The models suggest that 
galaxies belonging to region II have experienced numerous 
(i.e., $n_{b}\sim 3-15$) and extended ($\tau_{B}\simeq 60-100$ Myr) 
star formation events . 
We have found that a very good fit 
is provided by models for which the quiescent period between two 
successive bursts is of the same order as the gas infall 
timescale (i.e., $\tau_{f}\simeq \tau_{IB}$). This can be understood 
in the context of a hierarchical galaxy formation scenario where 
a major burst of star formation is activated each time a galaxy 
undergo a minor merger event, such as the accretion of a small 
satellite or primordial HI gas clouds. 

The evolutionary status of UV-selected galaxies, recently 
discussed in Contini et al. (2002), may be understood in the 
framework of these models. Their low N/O abundance ratios, 
with respect to other starburst galaxies with comparable 
metallicities, is well explained if they have just undergone 
a powerful starburst which enriched their ISM in oxygen. In 
these objects nitrogen may have not been completely released.  

\subsection{Metallicity-Luminosity relation}
\label{L_Z}

We now study the fundamental and well-known scaling relation between 
galaxy luminosity and metallicity (Contini et al. 2002 and references 
therein). This relation, which extends over $\sim 10$ 
magnitudes in luminosity and $\sim 2$ dex in metallicity 
seems to be an environmental (Vilchez 1995) and morphology-free 
(Aaronson 1986, Mateo 1998) relation. 

\begin{figure*}[!htb]
\includegraphics[clip=,angle=0,width=9cm]{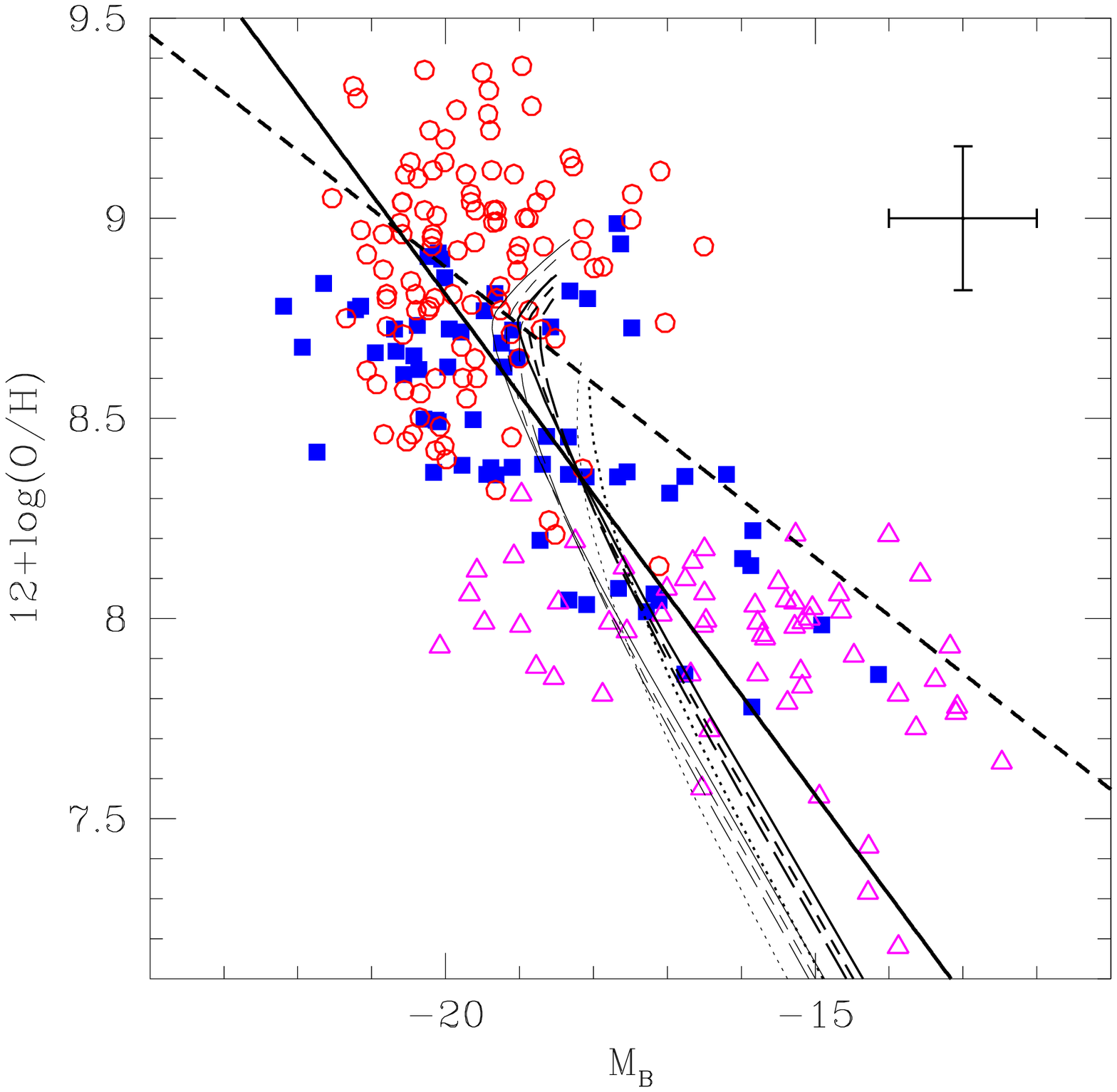}
\includegraphics[clip=,angle=0,width=9cm]{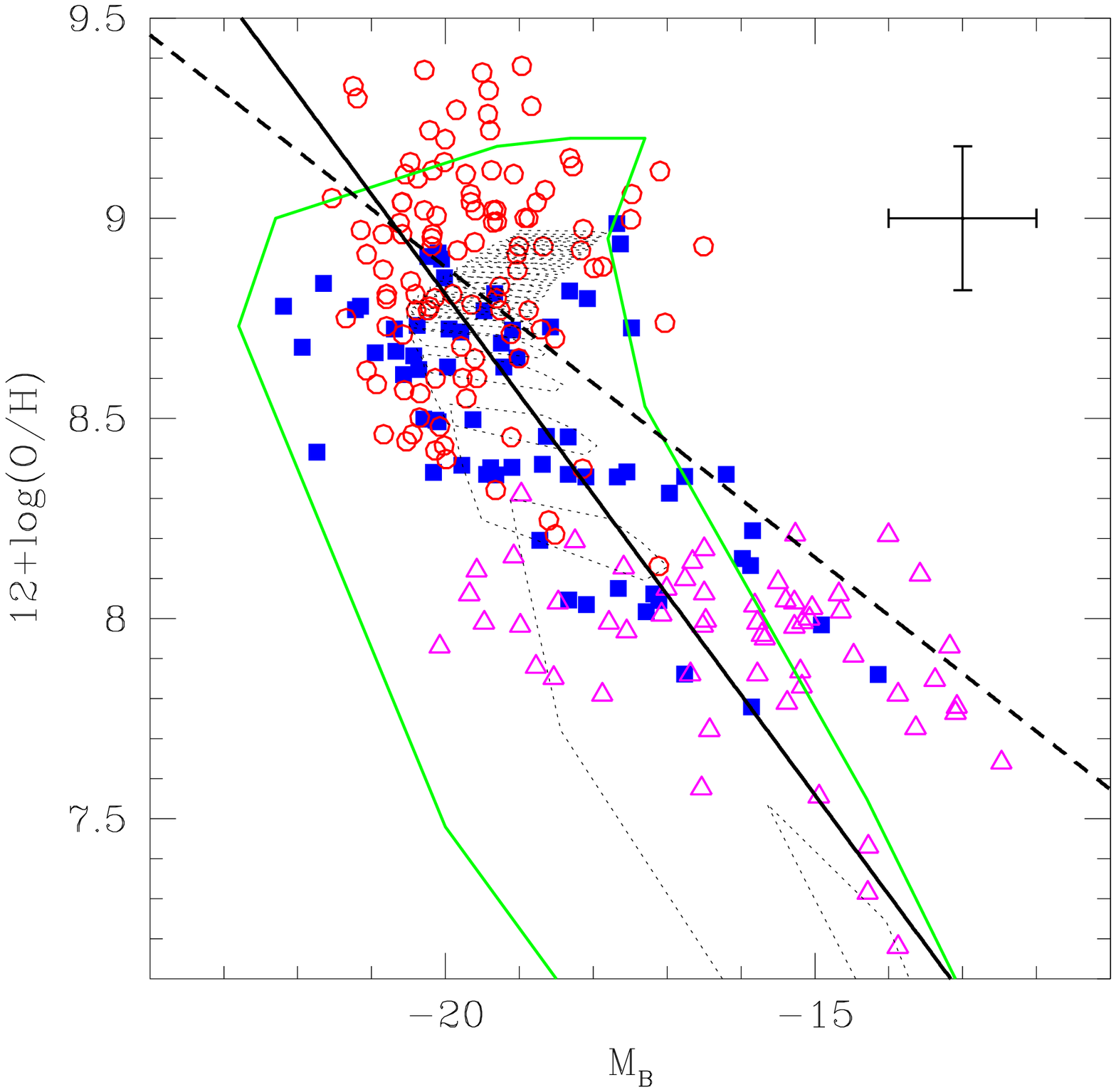}
\caption{The metallicity-luminosity relation for the starburst 
galaxy samples discussed in Section\,\ref{data}. Typical 
uncertainties are shown in the upper right of the right panel.
The solid thick line 
is a linear fit to the whole sample of starburst galaxies, and 
the dashed thick line is a linear fit to local {\it ``normal''} irregular 
and spiral galaxies (Kobulnicky \& Zaritsky 1999). 
The left panel shows model predictions assuming 
a continuous star formation scenario (see Fig.\,\ref{no_oh} for the 
legend). The right panel shows predictions assuming
a bursting star formation scenario. The model parameters are
the same as those used in Fig.\,\ref{no_oh}. The delineated area 
(thick line) encompasses all the model predictions calculated 
with the parameters reported in Table\,\ref{parameter}. For
illustration, the prediction of a specific model (dotted thin line) 
is shown.
}
\label{magB_metal}
\end{figure*} 

Richer \& McCall (1995) and Hunter \& Hoffman (1999) have 
reported that the dispersion in the metallicity of
dwarf galaxies of comparable luminosity increases with decreasing 
luminosity. In addition to significant uncertainties in the 
abundance determinations (Hidalgo-G\'amez \& Olofsson 1998), 
the high surface brightness of the star-forming regions in dwarf 
galaxies may cause a deviation from the mean relation 
(Roennback \& Bergvall 1995). Selective galactic 
winds and the fluctuations of the gas mass fraction among gas-rich 
dwarfs of a given luminosity (Pilyugin 2001) may also play a
significant role to produce the observed scatter for these galaxies. 
However, a significant dispersion is also observed for massive and 
bright galaxies (\mabs\ $\la -16$), for which the mechanisms invoked 
above are, in principle, not effective. Hence, one needs to come up 
with other mechanism(s) to account for the observed scatter in 
the latter galaxies.

Using the chemo-photometric evolution models discussed in the 
previous section, we now examine how much scatter 
affecting the metallicity-luminosity relation can be predicted. 
Note that proceeding in such a way to 
reproduce the metallicity-luminosity relation and its associated 
scatter, means implicitly that the relation under study is caused 
by the ability of galaxies to produce metals, rather than their 
ability to keep the products of their own evolution. 
In the left panel of Figure\,\ref{magB_metal}, we show the 
evolution of the metallicity as a function of the $B$-band absolute 
magnitude assuming continuous star formation.  
The model parameters and data samples are those discussed 
in the previous section (see Fig.\,\ref{no_oh}--left panel). 
Also shown are i) the linear fit to the whole sample of starburst 
galaxies (thick solid line) given by the following relation: 
$12+\log(O/H)=-0.25(\pm 0.01)\times M_{B}+3.81(\pm0.2)$, 
and ii) a linear fit to the {\it normal} irregular and spiral 
galaxies (Kobulnicky \& Zaritsky 1999). 

The left panel of Figure\,\ref{magB_metal} shows that models assuming 
continuous star formation are able to reproduce the ``mean'' relation 
between galaxy metallicity and luminosity. 
However, this plot shows that these models delineate a narrow 
region in the O/H vs. \mabs\ diagram, even if the free 
parameters were taken to span a quite large range of values. 
Hence, the plot tells us that continuous star formation models 
are unable to account for the observed scatter affecting the 
metallicity-luminosity relation.

In the right panel of Figure\,\ref{magB_metal}, we show the 
evolution of metallicity as a function of the $B$-band absolute 
magnitude, assuming that the star formation proceeds in successive 
bursts. The delineated area (thick line) in the O/H vs. 
\mabs\ plane encompasses all the model predictions calculated 
with the parameters reported in Table\,\ref{parameter}.
For illustration, the prediction of a specific model (dotted thin line) 
is plotted in more details. When the star formation is active, the 
$B$-band magnitude and the oxygen abundance increase simultaneously, 
as a consequence of the presence of hot massive stars, responsible for 
the major fraction of the $B$-band light, and of supernova explosions which 
enrich the star-forming region in heavy elements. 
During this phase the galaxy evolves along the left-hand boundary 
of the allowed area. During the interburst period, the $B$-band 
magnitude and the oxygen abundance decrease due to the star formation 
inactivity and to the dilution of the interstellar gas by the
metal-free gas infall. During this period, the galaxy evolves along 
the right-hand boundary of the allowed area. 
As the galaxy evolves, each starburst event consumes gas with enhanced 
element abundances in comparison to previous starburst events, leading 
to the intermittence between the two boundaries of the delineated
area.

Figure\,\ref{magB_metal} shows that the extent of the area allowed 
by intermittent star formation models is comparable to the 
observed spread of metallicity among galaxies of comparable 
luminosity.
More appropriate constraints on how the star formation proceeds 
in starburst galaxies, and which are their basic scaling relations 
may be derived using the N/H vs. $M_{\rm K}$ relation rather than 
O/H vs. \mabs. Indeed, nitrogen is less sensitive to the recent 
star formation history of galaxies than oxygen.
The same is true for the magnitude as the $B$-band is heavily 
affected by the most recent burst of star formation, and thus 
depend strongly on the evolutionary state of the galaxy. Near-infrared 
magnitudes will be more appropriate as a luminosity/mass indicator 
since it traces the total stellar mass of a galaxy 
(Contini, Mouhcine \& Davoust, in preparation). 

\section{Summary and conclusions}
\label{concl} 

We developed a chemical evolution model aimed at constraining 
the star formation history of starburst galaxies by investigating 
the origin of the dispersion observed in the evolution of 
both the N/O abundance ratio and galaxy luminosity as a function of 
metallicity. One of the main goals was to determine the 
range of values for the relevant model parameters characterizing the 
star formation history of starburst galaxies as a whole, i.e. over 
a large range of mass and metallicity.

The results of this investigation can be summarized as follows.
The comparison of model predictions with the available data samples 
on the abundance ratios of heavy elements such as O and N allows us 
to conclude that continuous star formation models can reproduce 
the scatter in the N/O versus O/H relation observed for 
low-mass \hii\ galaxies as a consequence of the systematic variation of 
the star formation efficiency among galaxies. 
However, continous star formation models are unable 
to reproduce i) the scatter observed for massive starburst and UV-selected 
galaxies in the N/O versus O/H relation and ii) the scatter in the 
\mabs\ versus O/H scaling relation observed for the whole sample 
of starburst galaxies. 

The dispersion associated to the distribution of N/O as a function 
of metallicity is well explained in the framework of bursting 
star formation models. It is interpreted as a consequence of 
the time-delay between the ejection of nitrogen and that of oxygen 
into the ISM. We found that metal-rich galaxies differ from 
metal-poor ones by their star formation efficiency and starburst
frequency. Indeed to match the observed 
mean evolution of N/O with metallicity and its dispersion for 
galaxies with $12+\log\,(O/H)\,\le\,8.5$, the models 
suggest that their star formation efficiency need to be low 
(between 0.5 and 5\,Gyr$^{-1}$), with a small number of star formation events.
On the contrary, the observed spread for galaxies located in the area 
where $12+\log\,(O/H)\,\ga\,8.5$, is best reproduced with models
assuming a higher star formation efficiency (between 7 and
25\,Gyr$^{-1}$), and more starburst events than metal-poor galaxies. 
We note also that only extended bursts ($\sim 50-100$ Myr) are compatible 
with the properties of massive metal-rich galaxies.

Bursting star formation models accounting for the observed scatter in 
the N/O vs. O/H diagram delineate an area in \mabs\ vs. O/H 
diagram where galaxies need to fall. Good agreement is found 
with the location of data in this diagram.

\begin{acknowledgements}
We thank R. Coziol and E. Davoust for useful suggestions.
\end{acknowledgements}

\end{document}